\documentclass[fleqn,12pt]{article}

\newtheorem{theo}{ }[section]
\newcommand{\marca}{\hspace{120mm} \= \kill}
\newcommand{\dis}{\displaystyle}
\newcounter{cuenta}

\begin{document}

\title{The Classification of Time Invariants (First Integrals) of Multinomial 
Systems of O.D.E.s and the Surprising Link Between Algebraic and 
Logarithmic Time Invariants, Dictated by the Method of Arrays.}

\author{Lawrence Goldman \thanks{lgoldman1@msn.com}}

\maketitle
 
\begin{abstract}

For a large class of systems of o.d.e.'s which have first integrals, the method of arrays yields the following results:
\begin{list}{\roman{cuenta})}{\usecounter{cuenta}}
\item The first integrals $I$ can be found by solving systems of linear equations.
\item How the first integral $I$ responds to changes in the system $S$.
\item An easy way for finding the first integral for a special class of first integrals if they exist.
\end{list}
\end{abstract}

\section{Introduction.}

Let $S$ be an autonomous system of o.d.e.'s of the form:

\begin{tabbing}
\marca
$ \dis S:\;\;y'_i=y_i\sum^r_{j=1}c_{ij}y_1^{h_{j1}}\ldots y_n^{h_{jn}}\;\;\;\;\;(i=1,\ldots,n) $ \> \parbox{15mm}{\begin{theo}\label{1.1}\end{theo}} \\
\end{tabbing}

In the study of the problem: how a trajectory determined by $S$, responds to changes in the system $S$; it becomes apparent, that limiting our attention to trajectories that are given by first integrals of the form:
\begin{tabbing}
\marca
$ \dis I=\sum^{q}_{k=1}e_k(y_1^{b_{k1}}\ldots y_n^{b_{kn}}), $
\> \parbox{15mm}{\begin{theo}\label{1.2}\end{theo}} \\
\end{tabbing}
the problem can be simplified. For, we can focus on the effect that changes in the exponent vectors $H_j=(h_{j1},\ldots,h_{jn})$ and or the coefficient vectors $C_j=(c_{1j},\ldots,c_{n_j})^t$, $(j=1,\ldots,r)$, of the system $S$, have on the coefficients $e_k$ and the exponent vectors $B_k=(b_{k1},\ldots,b_{kn})$ $(k=1,\ldots,q)$ of the first integral $I$.\\

This was done in my paper: Integrals of Multinomial Systems of Ordinary Differential Equations (Journal of Pure and Applied Algebra, 45 (1987) 225-240).\\

To facilitate the computation, a more efficient notation was employed. In this notation the system $S$ of (\ref{1.1}) becomes:
\begin{tabbing}
\marca
$ \dis S:\;\;y'=y\sum^r_{j=1}C_jY^{H_j},\mbox{ where}$
\> \parbox{15mm}{\begin{theo}\label{1.3}\end{theo}} \\
\end{tabbing}
\vspace{-3mm}\begin{quote}
 $y$ is the column vector $(y_1,\ldots,y_n)^t$,

 $C_j$ is the column vector $(c_{1j},\ldots,c_{n_j})^t$ and

 $H_j$ is the row vector $(h_{j1},\ldots,h_{jn})$.
\end{quote}

Similarly, (\ref{1.2}) becomes:
\begin{tabbing}
\marca
$ \dis I=\sum_{k=1}^{q}e_kY^{B_k},$
\> \parbox{15mm}{\begin{theo}\label{1.4}\end{theo}} \\
\end{tabbing}
where $B_k$ is the row vector $(b_{k1},\ldots,b_{kn})$.\\
We will refer to (\ref{1.3}) and (\ref{1.4}) as the multinomial vector form of $S$ and $I$ (m.v.f. hereafter).

In this notation, the formula for the derivative of a monomial $Y^B$, along the trajectory, is easily shown to be:
\begin{tabbing}
\marca
$ \dis (Y^B)'=Y^B\sum^r_{j=1}(B;C_j)Y^{H_j},$
\> \parbox{15mm}{\begin{theo}\label{1.5}\end{theo}} \\
\end{tabbing}
where $(B;C_j)=\sum^n_{i=1}b_ic_{ij}$ (the inner product of the vectors $B$ and $C_j$).Using (\ref{1.5}), the derivative of $I$ in (\ref{1.4}) becomes:
\[I'=\sum_{k=1}^{q}e_k\sum_{j=1}^r(B_k;C_j)Y^{B_k+H_j}=0,\]
neglecting all terms for which $(B_k;C_j)=0$ and grouping the coefficients of equal monomials together, we get:
\begin{tabbing}
\marca
$ \dis I'=\sum_{i=1}^p(\sum_{H_{\alpha}+B_k=E_i}e_k(B_k;C_{\alpha}))Y^{E_i}=0,$
\> \parbox{15mm}{\begin{theo}\label{1.6}\end{theo}} \\
\end{tabbing}
from which the following was proven:
\begin{tabbing}
\hspace{5mm} \= \hspace{115mm} \= \kill
\\
 \> \parbox{110mm}{a) Each $B_k\;\;(k=1,\ldots,q)$ satisfies a system of $r$ linear equations. \\
\\
  b) The difference of any two exponent vectors of $I$ is a linear combination of exponent vectors of $S$.\\
\\
c) $e_1,\ldots,e_q$ is a solution of $p$ linear homogeneous equations (the equation are the coefficients of $Y^{E_i}$) which must vanish, since $y_1,\ldots,y_n$ are independent.} \> \parbox{15mm}{\begin{theo}\label{1.7}\end{theo}} \\
\\
\end{tabbing}

To make it possible to classify first integrals by the various relations between $S$ and $I$ implied by (\ref{1.7}), the 'method of arrays' was developed. An integral array corresponding to a given first integral $I$ of a system $S$, is a pictorial representation of all the conditions required for $I$ to be a first integral of $S$.

In this paper we derive similar results for the cases where $I$ is given by:
\begin{tabbing}
\marca
$ \dis I=\ln\left(Y^{B_1}\right)+\sum_{k=2}^{q}e_kY^{B_k}$ \> \parbox{15mm}{\begin{theo}\label{1.8}\end{theo}} \\
\end{tabbing}
\begin{tabbing}
\marca
$ \dis I=e_1Y^{B_1}+\ln\left(1+\sum_{k=2}^{q}e_kY^{B_k}\right),$ \> \parbox{15mm}{\begin{theo}\label{1.9}\end{theo}} \\
\end{tabbing}
with some surprising results. In the case of (\ref{1.8}), we find that the logarithmic integral is closely linked to an algebraic integral of the form (\ref{1.4}), in the following sense. Let $I$, given by (\ref{1.4}), be a first integral of a system $S$ given by (\ref{1.3}), where the coefficients and exponents of $S$ are real or complex numbers. We will show the existence of a multinomial system $S(\theta)$, depending on a set of parameters $\theta=\theta_1,\ldots,\theta_m$ satisfying the following:\\

\begin{tabbing}
\hspace{5mm} \= \hspace{115mm} \= \kill
 \> \parbox{110mm}{a) $S(\theta)$ has a first integral $I(\theta)$ such that $e_k(\theta)$ and the components of $B_k(\theta)\;\;(k=1,\ldots,q)$ are rational functions of $\theta_1,\ldots,\theta_m$. \\
\\
 b) There exist $\vec{\theta}$ such that $S(\vec{\theta})=S,\;I(\vec{\theta})=I$  \\
\\
c) If for some value $\theta^{\ast},\;I(\theta^{\ast})$ reduces to a non-zero constant, then there exists $I^{\ast}(\theta^{\ast})=\ln\left(Y^{B_1(\theta^{\ast})}\right)+\sum_{k=2}^qe_kY^{B_k(\theta^{\ast})}$ which is a first integral of $S(\theta^{\ast})$. Thus $S(\vec{\theta})$ which has an algebraic first integral given by (\ref{1.4}) and $S(\theta^{\ast})$ which has a logarithmic first integral given by (\ref{1.8}), both belong to the continuous system $S(\theta)$} \> \parbox{17mm}{\begin{theo}\label{1.10}\end{theo}} \\
\\
\end{tabbing}
In the case where the first integral of $S$ is of the form (\ref{1.9}) there exists $S(\theta,\rho)$ where $\theta$ is a set of continuous parameters as above, while $\rho$ takes on positive integral values only. $S(\theta,\rho)$ has a first integral given by 
\begin{tabbing}
\marca
$ \dis I(\theta,\rho)=e_1(\theta,\rho)Y^{B_1(\theta,\rho)}+\ln\left(1+\sum_{k=2}^{\rho}e_k(\theta,\rho)Y^{B_k(\theta,\rho)}\right)$ \> \parbox{17mm}{\begin{theo}\label{1.11}\end{theo}} \\
\end{tabbing}
where $e_k(\theta,\rho),B_k(\theta,\rho)$ are rational functions of $\theta_1,\ldots,\theta_m,\rho$ and there exist $\vec{\theta},\vec{\rho}=q$ s.t. $S(\vec{\theta},\vec{\rho})=S,\;I(\vec{\theta},\vec{\rho})=I$. Note that $\rho$ is the number of monomials in $I(\theta,\rho)$.

In addition, the method of arrays yields some curious results, such as:

\begin{list}{\roman{cuenta})}{\usecounter{cuenta}}
\item  Let $S$ be given by
\[ y^{(n)}=f\left(y,y^{\prime},\ldots,y^{(n-1)}\right)=\sum_{j=1}^s l_j Y^{M_j} \]

where \( \dis Y^{M_j}=\prod_{i=0}^{n-1}\left(y^{(i)}\right)^{m_{ji}}, j=1,\ldots,s.\)

Let $S$ have a first integral

\[I=\sum_{k=1}^qe_k Y^{B_k}, Y^{B_k}=\prod_{i=0}^{n-1}\left(y^{(i)}\right)^{b_{kxhosti}} \]

then, the exponents of $I, B_K (k=1,\ldots,q)$ are independent of $l_j (j=1, \ldots, s)$ provided:

\begin{tabbing}
\marca
$ \dis Y^{M_j}\neq \left(y^{(i)}\right)^{-1}y^{(i+1)}y^{(n-1)}$ \>\parbox{17mm}{\begin{theo}\label{1.12}\end{theo}}
\end{tabbing}
e.g. Let $S$ be given by

\[ y^{\prime \prime}=-l_1 y^{-1}\left(y^{\prime}\right)^2+l_2 y+l_3 y^3 \]

$Y^{M_1}$ violates (\ref{1.12}), while $Y^{M_2}, Y^{M_3}$ do not. $S$ has a first integral

\[ I=(l_1+1) y^{2 l_1} (y^{\prime})^2 - l_2 y^{2 (l_1+1)}-\frac{l_3 (l_1+1) y^{2 (l_1+2)}}{l_1+2} \]

$l_1$, the coefficient of $Y^{M_1}$ appears in every exponent of $I$, while $l_2,l_3$ appear in none.

\item Let $S$ be given by \\

\( \dis S : y^{\prime}=y\sum_{j=1}^rC_jY^{H_j}, \) let $S$ have a first integral which is either:\\

\[ \mbox{a) } I=\sum_{k=1}^qe_kY^{B_k} \mbox{  or} \]

\[ \mbox{b) } I=\ln\left(Y^{B_1}\right)+\sum_{k=2}^qe_kY^{B_k} \]

Let $\sigma_{\alpha}(S)$ denote the system:

\[ y^{\prime}=y\sum_{j=1}^rC_jY^{\alpha H_j},\;\; \alpha \mbox{ any real or complex number.}\]

Then $\sigma_{\alpha}(S)$ has a first integral:

\[ \sigma_{\alpha}(I)=\sum_{k=1}^q\bar{e_k}Y^{\bar{B_k}}, \] where $\bar{B_k}=\alpha B_k, \;\; \bar{e_k}=e_k \; (k=1,\ldots,q).$

\end{list}

\section{The Derivative Formulas.}
\vspace{14pt}
 Throughout this paper we assume that $y_1,\ldots,y_n$ are algebraically independent.
Let $y=(y_1,\ldots,y_n)$ be a solution of the system $S$ given by (\ref{1.1}) and let $Y^B=y_1^{b_1}\ldots y_n^{b_n}$,
\[ (Y^B)'=\sum_{i=1}^nb_iy^{-1}_iY^By'_i=Y^B\sum_{j=1}^r\sum_{i=1}^n(b_ic_{ij})Y^{H_j}\;\;\;\mbox{yielding} \]
\begin{tabbing}
\marca
$ \dis \left\{ \begin{array}{ll}
i)\;\;\;(Y^B)'=Y^B\sum_{j=1}^r(B;C_j)Y^{H_j}\\
ii)\;\;\;\left( \ln\left(Y^B\right) \right)'=Y^{-B}(Y^B)'=\sum_{j=1}^r(B;C_j)Y^{H_j}
\end{array}\right.$ \> \parbox{15mm}{\begin{theo}\label{2.1}\end{theo}} \\
\end{tabbing}
$i)$ of (\ref{2.1}) has two interesting consequences:
\begin{tabbing}
\marca
\parbox{120mm}{\begin{description}
\item a) A necessary and sufficient condition for a monomial $Y^B$ to be a first integral of $S$ is that $(B;C_j)=0\;\;(j=1,\ldots,n)$, this implies:
\subitem $i)$ monomial first integrals of $S$ are independent of the exponent vectors $H_j$ of $S\;\;(j=1,\ldots,r)$.
\subitem $ii)$ Let the matrix $(c_{ij})$ of $S$ have rank $s\leq n$, then $S$ has exactly $n-s$ independent first integrals. For the system of homogeneous linear equations:\\
$(B;C_j)=0\;\;\;\;\;(j=1,\ldots,r)$\\
has exactly $n-s$ independent solutions.
\item b) Let $r=n$ and let $(H_i,C_j)=0$ for all $i\neq j$, then \\
$(Y^{H_i})'=(H_i,C_i)Y^{2H_i}\;\;\;\;(i=1,\ldots,n).$ \\
Setting $z_i=Y^{H_i}$ yields a complete separation of variables, in $z_1,\ldots,z_n$.
\end{description}}\> \parbox{15mm}{\begin{theo}\label{2.2}\end{theo}} \\
\end{tabbing}
Remark: The sufficiency condition of a) applies to any system of the form:
\[ y'_i=y_i\sum_{j=1}^rc_{ij}f_j(y_1,\ldots,y_n).\]
b) applies to any system of the form:
\[ y'_i=y_i\sum_{j=1}^rc_{ij}f_j(Y^{H_j}). \]

\section{The Method of Arrays.}

Before we give a systematic treatment of the ``method of arrays'', we give a few examples to show how to construct the continuous system $S(\theta)$ and its first integral $I(\theta)$, to which a given system $S$ and its first integral $I$ belong.

This can be done for any system $S$ of the form (\ref{1.1}) whose first integral $I$ is such that\\
\\
$ \dis I'=\frac{A(y)}{B(y)}\;\;\mbox{ where }A(y),B(y)$ \\
\\
are linear combinations of monomials.\\

Example 1.
\begin{tabbing}
\marca
$\left\{ \begin{array}{ll} 
y'_1=y_2=y_1(y_1^{-1}y_2) & \mbox{ or }\;y'=y(C_1Y^{H_1}+C_2Y^{H_2}) \\
y'_2=-y_1=y_2(-y_1y_2^{-1})
\end{array}\right.$ \> \parbox{15mm}{\begin{theo}\label{3.1}\end{theo}}
\end{tabbing}
where,
\[ H_1=(-1,1),\;\;H_2=(1,-1),\;\;C_1=\left(\begin{array}{c} 1 \\ 0 \end{array}\right),\;C_2=\left(\begin{array}{c} 0 \\ -1 \end{array}\right).\] 
(the latter is the m.v.f. of $S$). S has a first integral:
\begin{tabbing}
\marca
\parbox{110mm}{$I=y^2_2+y_1^2=e_1Y^{B_1}+e_2Y^{B_2},$\\
where $\;B_1=(0,2),\;B_2=(2,0),\;e_1=e_2=1$}
\> \parbox{15mm}{\begin{theo}\label{3.2}\end{theo}}\\
\end{tabbing}

Applying the derivative formula $i)$ of (\ref{2.1}) to $I$, we get:
\begin{tabbing}
\marca
$\left\{ \begin{array}{lcl}
I' & = & e_1(Y^B_1)'+e_2(Y^{B_2})'\\
   & = & e_1[(B_1;C_1)Y^{B_1+H_1}+(B_1;C_2)Y^{B_1+H_2}] \\
   & + & e_2[(B_2;C_1)Y^{B_2+H_1}+(B_2;C_2)Y^{B_2+H_2}]=0.
\end{array}\right.$ \> \parbox{15mm}{\begin{theo}\label{3.3}\end{theo}} \\
\end{tabbing}

Substituting for $B_1,B_2,C_1,C_2,H_1,H_2$ their values, given by (\ref{3.1}), (\ref{3.2}), in (\ref{3.3}) we find:\\
\\
$ (B_1;C_1)=(B_2,C_2)=0 $\\
\\
$ B_1+H_2=B_2+H_1=E_1,\;\;\;\mbox{ so that} $\\
\\
$ I'=[e_1(B_1;C_2)+e_2(B_2;C_1)]Y^{E_1}=0 $ \\

Thus, the relations between $I$ and $S$ are:
\begin{tabbing}
\marca
$\left\{ \begin{array}{l}
i)\;\;\;(B_1;C_1)=(B_2;C_2)=0 \\
ii)\;\;\;B_1+H_2=B_2+H_1=E_1 \\
iii)\;\;\;e_1(B_1;C_2)+e_2(B_2;C_1)=0
\end{array}\right.$ \> \parbox{15mm}{\begin{theo}\label{3.4}\end{theo}} \\
\end{tabbing}
These relations are summarized by the array:
\[ {\mathbf A}=
\begin{array}{ccc}
B_1 & B_2 & \\
(H_2 & H_1) & E_1 
\end{array} \]

Which obeys the following:
\begin{tabbing}
\marca
$\left\{ \begin{array}{l} 
\parbox{110mm}{$ i)\;\;H_{\alpha}$ appears in the $k^{\mbox{\small th}}$ column of a $p\times q$ array if and only if $(B_k,C_{\alpha})\neq0.$ \\
$ ii)\;\;H_{\alpha}$ appears in the $j^{\mbox{\small th}}$ row and the $k^{\mbox{\small th}}$ column of a $ p\times q$ array if and only if: $ B_k+H_{\alpha}=E_j,\;\;1\leq j\leq p,\;\;1\leq k\leq q.$}
\end{array}\right.$ \> \parbox{15mm}{\begin{theo}\label{3.5}\end{theo}}\\
\end{tabbing}

In our case the array tells us that:
\[ I'=[e_1(B_1;C_2)+e_2(B_2;C_1)]Y^{E_1}=0 \]
where 
\[ E_1=B_1+H_2=B_2+H_1. \]

Now, $ii)$ of (\ref{3.4}) and $i)$ of (\ref{3.5}) imply:
\begin{tabbing}
\marca
$\left\{ \begin{array}{lcl}
(B_1;C_2) & = & (B_1-B_2;C_2)+(B_2;C_2)\\
          & = & (H_1-H_2;C_2)+0=(H_1-H_2;C_2)\neq0\\
(B_2;C_1) & = & (B_2-B_1;C_1)+(B_1;C_1)\\
          & = & (H_2-H_1;C_1)+0=(H_2-H_1;C_1)\neq0
\end{array}\right.$ \> \parbox{15mm}{\begin{theo}\label{3.6}\end{theo}} \\
\end{tabbing}

We can, now, use $i)$ of (\ref{3.4}) and (\ref{3.6}) to solve for $B_1, B_2,e_1,e_2$. For, $B_1,B_2$ are solutions of the linear systems:
\begin{tabbing}
\marca
$i)\;\;\left\{ \begin{array}{ll}
(B_1;C_1)=0\\
(B_1;C_2=(H_1-H_2;C_2)
\end{array}\right. \hspace{10mm}
ii)\left\{ \begin{array}{ll}
(B_2;C_2)=0\\
(B_2;C_1)=(H_2-H_1;C_1).
\end{array}\right.$ \> \parbox{15mm}{\begin{theo}\label{3.7}\end{theo}} \\
\end{tabbing}
 
These systems have unique non-zero solutions provided:
\begin{tabbing}
\marca
$\left\{ \begin{array}{l}
i)\;\;\;d=\det \left( \begin{array}{ll} c_{11} & c_{12}\\
                                  c_{21} & c_{22}\end{array}\right) \neq0\\
ii)\;\;\;(H_1-H_2;C_1)\neq0\\
iii)\;\;\;(H_1-H_2;C_2)\neq0.
\end{array}\right.$ \> \parbox{15mm}{\begin{theo}\label{3.8}\end{theo}}\\
\end{tabbing}
 
We can now use $iii)$ of (\ref{3.4}) to find $e_1,e_2$.

Since the only conditions on $S$ are the inequalities (\ref{3.8}), we may take for $S(\theta)$, the continuous system to which $S$, given by (\ref{3.1}), belongs, the full 8 parameter system:
\begin{tabbing}
\marca
$ S(\theta):\;\;\;y'=y[\left( \begin{array}{c} c_{11}\\ c_{21}\end{array}\right) Y^{(h_{11},h_{12})}+\left( \begin{array}{c} c_{12}\\c_{22}\end{array}\right) Y^{(h_{21},h_{22})}] $ \> \parbox{15mm}{\begin{theo}\label{3.9}\end{theo}} \\
\end{tabbing}
subject only to the inequalities (\ref{3.8}).

We can, now, solve the linear systems (\ref{3.7}) for $B_1(\theta),B_2(\theta)$, and obtain:
\begin{tabbing}
\marca
$\left\{ \begin{array}{l}
i)\;\;B_1(\theta)=\frac{(H_1-H_2;C_2)}{d}(-c_{21},c_{11})\\
ii)\;\;B_2(\theta)=\frac{(H_2-H_1;C_1)}{d}(c_{22},-c_{12}) \end{array}\right. $ \> \parbox{17mm}{\begin{theo}\label{3.10}\end{theo}} \\
\end{tabbing}

To check that (\ref{3.10}) are solutions to (\ref{3.7}), note that $(-c_{21},c_{11}),(c_{22},-c_{12})$ are normal to $C_1,C_2$ respectively and that
\[ ((-c_{21},c_{11});C_2) =d\;\;\;\;\;
((c_{22},-c_{12});C_1) =d. \]

$B_2(\theta)$ may also be obtained, when $B_1(\theta)$ is known, by using $ii)$ of (\ref{3.4}) yielding:
\begin{tabbing}
\marca
$ B_2(\theta)=B_1(\theta)+H_2-H_1 $ \> \parbox{17mm}{\begin{theo}\label{3.11}\end{theo}} \\
\end{tabbing}

To show that (\ref{3.11}) agrees with $ii)$ of (\ref{3.10}), we show that (\ref{3.11}) is a solution of the linear system:
\[ (B_2(\theta);C_2)=0 \]
\[ (B_2(\theta);C_1)=(H_2-H_1;C_1) \]
 
 For \[ \begin{array}{lcl} (B_2(\theta);C_2) & = & (B_1(\theta);C_2)+(H_2-H_1;C_2) \\
                                           &  = & (H_1-H_2;C_2)+(H_2-H_1;C_2)=0  \\
                (B_2(\theta),C_1) & = & (B_1(\theta),C_1)+(H_2-H_1,C_1)=(H_2-H_1,C_1),
 \end{array} \]

Thus (\ref{3.11}) and $ii)$ of (\ref{3.10}) are both solutions of the linear system $ii)$ of (\ref{3.7}) which has a unique solution when the determinant $d$ is not equal to zero.

To find $e_1(\theta),e_2(\theta)$ we use $iii)$ of (\ref{3.4})
\[ e_1(H_1-H_2;C_2)+e_2(H_2-H_1,C_1)=0,\;\;\;\mbox{ yielding} \]
\[ e_1=(H_2-H_1;C_1),\;\;e_2=(H_2-H_1,C_2) \]
Thus
\begin{tabbing}
\marca
$ \begin{array}{lll}
I(\theta) & = & (H_2-H_1,C_1)(y_1^{-c_{21}}y_2^{c_{11}})^{\frac{(H_1-H_2;C_2)}{d}} \\
          &   &                                                                    \\
          & + &(H_2-H_1,C_2)(y_1^{c_{22}}y_2^{-c_{12}})^{\frac{(H_2-H_1,C_1)}{d}}  \\
\end{array} $ \> \parbox{17mm}{\begin{theo}\label{3.12}\end{theo}} \\
\end{tabbing}

setting $\bar{c}_{12}=\bar{c}_{21}=0,\;\bar{c}_{11}=1,\;\bar{c}_{22}=-1$ we get $\bar{d}=-1$ and $H_1=(-1,1),\;H_2=(1,-1)$ in (\ref{3.8}) and (\ref{3.12}) we get:
\[ S(\bar{\theta})=S\;\;\mbox{ of (\ref{3.1}) }\;\;\mbox{ and }I(\bar{\theta})=2I\mbox{ of (\ref{3.2})} \]

We now show that the system given by (\ref{3.9}) has a first integral even when the inequalities are violated (one at a time).

Let $d=0$ then $C_2=lC_1$ and by $i)$ of (2.2) $S$ has the monomial first integral $Y^B$ where $(B;C_1)=(B;C_2)=0$.
Now, let $d\neq0$ and let $\theta^{\ast}$ be such that $(H_1-H_2,C_2)=0$ while $(H_1-H_2,C_1)\neq0$, then

$I(\theta^{\ast})=(H_2-H_1;C_1)$ and fails to define a first integral of $S(\theta^{\ast})$. Fortunately, $(H_1-H_2,C_2)=0$ is the very condition required for $S(\theta^{\ast})$ to have a first integral of the form:
\begin{tabbing} 
\marca
$ \dis I^{\ast}(\theta^{\ast})=\ln\left(Y^{B_1}\right)+Y^{B_2} $ \> \parbox{17mm}{\begin{theo}\label{3.13}\end{theo}} \\
\end{tabbing}
For applying the derivative formula $ii)$ of (\ref{2.1}) to $I^{\ast}(\theta^{\ast})$ we get
\begin{tabbing}
\marca
$ \dis (I^{\ast}(\theta^{\ast}))'=(B_1;C_2)Y^{H_2}+(B_2;C_1)Y^{B_2+H_1}=0 $ \> \parbox{17mm}{\begin{theo}\label{3.14}\end{theo}} \\
\end{tabbing}
Thus the relations between $S(\theta^{\ast})$ and $I^{\ast}(\theta^{\ast})$ are:
\begin{tabbing}
\marca
$ \begin{array}{l}
i)\;\;\;(B_1;C_1)=(B_2;C_2)=0 \\
ii)\;\;\;H_2=B_2+H_1=E_1 \\
iii)\;\;\;(B_1;C_2)+(B_2;C_1)=0 \\
\end{array} $ \> \parbox{17mm}{\begin{theo}\label{3.15}\end{theo}} \\
\end{tabbing}
$ii)$ of (\ref{3.15}) implies $B_2=H_2-H_1$ and $i)$ and $iii)$ of (\ref{3.15}) yield the system of linear equations:
\begin{tabbing}
\marca
$ \dis \left\{ \begin{array}{l}
(B_1,C_1)=0 \\
(B_1,C_2)=-(B_2,C_1)=(H_1-H_2;C_1) \\
\end{array}\right.$ \> \parbox{17mm}{\begin{theo}\label{3.16}\end{theo}} \\
\end{tabbing}
\[ \mbox{and }\;\;\;B_1=\frac{(H_1-H_2;C_1)}{d}(-c_{21},c_{11}) \]
\begin{tabbing}
\marca
$ \dis I^{\ast}(\theta^{\ast})=\frac{(H_1-H_2;C_1)}{d}\ln\left(y_1^{-c_{21}}y_2^{c_{22}}\right)+y_1^{h_{21}-h_{11}}y_2^{h_{22}-h_{12}} $ \> \parbox{17mm}{\begin{theo}\label{3.17}\end{theo}} \\
\end{tabbing}
The integral array of $I^{\ast}(\theta^{\ast})$ is ${\mathbf A}^{\ast}=(\stackrel{o}{H}_2H_1)$ (the $\circ$ above $H_2$ indicates that $B_1$ is not added to $H_2$ to get $ii)$ of (\ref{3.15})).
Similarly, if $S(\theta^{\ast\ast})$ is such that $(H_1-H_2,C_2)\neq0,\;d\neq0$ but, $(H_1-H_2;C_1)=0$ then the relations between $I^{\ast}(\theta^{\ast\ast})$ and $S(\theta^{\ast\ast})$ are:
\begin{tabbing}
\marca
$ \begin{array}{l}
i)\;\;\;(B_1,C_1)=(B_2,C_2)=0 \\
ii)\;\;\;B_1+H_2=H_1=E_1 \\
iii)\;\;\;(B_1;C_2)+(B_2;C_1)=0 \\
\end{array}$ \> \parbox{17mm}{\begin{theo}\label{3.18}\end{theo}} \\
\end{tabbing}
the integral array of $I^{\ast}(\theta^{\ast\ast})$ is
\[ {\mathbf A}^{\ast\ast}=(H_2\;\stackrel{o}{H}_1)\;\;\;\;\mbox{ and} \]
\begin{tabbing}
\marca
$ \dis I^{\ast}(\theta^{\ast\ast})=y_1^{(h_{11}-h_{21})}y_2^{(h_{12}-h_{22})}+\frac{(H_2-H_1,C_2)}{d}\ln\left(y_1^{c_{22}}y_2^{-c_{12}}\right) $ \> \parbox{17mm}{\begin{theo}\label{3.19}\end{theo}} \\
\end{tabbing}

Summarizing the above:

The system $S$ of (\ref{3.1}) and its first integral $I$ of (\ref{3.2}) belong to the continuous 8 parameter system $S(\theta)$ of (\ref{3.9}) and its first integral $I(\theta)$ of (\ref{3.12}) which exists provided $S(\theta)$ satisfies the 3 inequalities:
\[ \begin{array}{l}
i)\;\;d\neq0               \\
ii)\;\;(H_1-H_2;C_1)\neq0  \\
iii)\;\;(H_1-H_2;C_2)\neq0 \\
\end{array} \]
If $d=0,\;S(\theta)$ has the monomial integral $I=Y^B$, where $(B;C_1)=(B;C_2)=0$.

If $S(\theta^{\ast})$ is such that $d\neq0,\;(H_1-H_2,C_1)\neq0$ but $(H_1-H_2;C_2)=0$, then $S(\theta^{\ast})$ has the logarithmic integral $I^{\ast}(\theta^{\ast})$ given by (\ref{3.17}).

If $d\neq0,\;(H_1-H_2;C_2)\neq0$ but $(H_1-H_2,C_1)=0$, when $\theta=\theta^{\ast\ast}$. Then $S(\theta^{\ast\ast})$ has the logarithmic integral $I^{\ast}(\theta^{\ast\ast})$ given by (\ref{3.19}).

\section{} 

Let $S_3(\theta)$ be the system given by:
\[ y'=y[C_1Y^{H_1}+C_2Y^{H_2}+C_3Y^{H_3}],\;\;\mbox{ where} \]
$C_1,H_1,C_2,H_2$ are as in (\ref{3.9}), subject only to the three inequalities (\ref{3.8}). We shall refer to the system $S(\theta)$ of (\ref{3.9}) as $S_2(\theta)$ and write
\[ S_3(\theta)=S_2(\theta)+C_3Y^{H_3} \]
Since $n=2$ and $C_1,C_2$ are linearly independent, we may write
\begin{tabbing}
\marca
$ \dis C_3=l_1C_1+l_2C_2 $ \> \parbox{15mm}{\begin{theo}\label{4.1}\end{theo}} \\
\end{tabbing}
Now, $S_2(\theta)$ has a first integral $I_2(\theta)$ given by (\ref{3.12}). We are going to show that $S_3(\theta)$ has a first integral
\[ I_3(\theta)=I_2(\theta)+e_3Y^{B_3} \]
subject only to the following:
\begin{tabbing}
\marca
$ \dis (H_3;C_3)=l_1(H_1;C_1)+l_2(H_2;C_2) $ \> \parbox{15mm}{\begin{theo}\label{4.2}\end{theo}} \\
\end{tabbing}
where $l_1,l_2$ are as in (\ref{4.1}), and $S_3(\theta)$ satisfies the following additional inequalities:
\begin{tabbing}
\marca
$ \dis \left\{\begin{array}{l}
i)\;\;\;(H_1-H_3;C_1,C_3)\neq0  \\
ii)\;\;\;(H_2-H_3;C_2,C_3)\neq0 \\
\end{array}\right.$ \> \parbox{15mm}{\begin{theo}\label{4.3}\end{theo}} \\
\end{tabbing}
There are, now, 2 cases to consider.

Case 1.
\begin{tabbing}
\marca
$ \dis l_1\times l_2\neq0 $ \> \parbox{15mm}{\begin{theo}\label{4.4}\end{theo}} \\
\end{tabbing}
This inequality and the inequality $d=d_{12}\neq0$ imply that any two of $C_1,\;C_2,\;C_3$ are linearly independent. To find $I_3(\theta)$, let
\begin{tabbing}
\marca
$ \dis B_3=B_1+H_3-H_1 $ \> \parbox{15mm}{\begin{theo}\label{4.5}\end{theo}} \\
\end{tabbing}
where $B_1$, is as in $i)$ of (\ref{3.10}). We show that (\ref{4.2}) implies $(B_3;C_3)=0$. For, using $i)$ of (\ref{3.7}) we get:
\begin{tabbing}
\marca
$ \dis \left\{ \begin{array}{lccl}
i)   & (B_3;C_3) & = & (B_1+H_3-H_1;C_3)                                  \\
     &           & = & (B_1;C_3)+(H_3,C_3)-(H_1,C_3)                      \\
     &           & = & (B_1;l_1C_1+l_2C_2)+l_1(H_1,C_1)                   \\
     &           & + & l_2(H_2;C_2)-l_1(H_1,C_1)-l_2(H_1,C_2)             \\
     &           & = & l_2(B_1,C_2)+l_2(H_2-H_1;C_2)                      \\
     &           & = & l_2[(H_1-H_2,C_2)+(H_2-H_1,C_2)]=0                 \\
     &\mbox{Also}&   &                                                    \\
ii)  & (B_3,C_1) & = & (B_1+H_3-H_1;C_1)                                  \\
     &           & = & (B_1;C_1)+(H_3-H_1;C_1)                            \\
     &           & = & 0+(H_3-H_1;C_1)\neq0                               \\   
iii) & (B_3;C_2) & = & (B_1+H_3-H_1;C_2)                                  \\
     &           & = & (B_1;C_2)+(H_3-H_1;C_2)                            \\
     &           & = & (H_1-H_2;C_2)+(H_3-H_1;C_2)                        \\
     &           & = & (H_3-H_2;C_2)\neq0                                 \\
iv)  & (B_1,C_3) & = & (B_3+H_1-H_3;C_3)                                  \\
     &           & = & (B_3,C_3)+(H_1-H_3;C_3)=(H_1-H_3;C_3)\neq0         \\   
v)   &  B_2+H_3  & = & B_1+H_2-H_1+H_3                                    \\ 
     &           & = & B_1+H_3-H_1+H_2=B_3+H_2                            \\
vi)  & (B_2;C_3) & = & (B_2-B_3;C_3)+(B_3;C_3)                            \\
     &           & = & (H_2-H_3,C_3)+0\neq0                               \\ 
\end{array}\right. $ \> \parbox{15mm}{\begin{theo}\label{4.6}\end{theo}}  \\
\end{tabbing}
To get the results of (\ref{4.6}) we used (\ref{3.4}), (\ref{4.2}), (\ref{4.3}), (\ref{4.4}) and (\ref{4.5}).

The relations (\ref{4.6}) are summarized by the array:
\begin{tabbing}
\marca
$ \dis  {\mathbf A}= \begin{array}{cc}
 \begin{array}{ccc}
    B_1    &    B_2    &   B_3    \\
 \end{array}                          &    \\
 \left( \begin{array}{ccc}
    H_2    &    H_1    & \mbox{\O} \\
    H_3    & \mbox{\O} &   H_1    \\
 \mbox{\O} &    H_3    &   H_2    \\
\end{array}\right)                    & \begin{array}{c} 
                                        E_1 \\ 
                                        E_2 \\ 
                                        E_3 \\
                                        \end{array}  
\end{array} $ \> \parbox{15mm}{\begin{theo}\label{4.7}\end{theo}} \\
\end{tabbing}
Columns $k$ says $(B_k;C_j)\neq0$ if and only if $k\neq j\;\;\;\;(k=1,2,3)$
Row 1 gives the relations (\ref{3.4}).\\
Row 2 implies $B_1+H_3=B_3+H_1=E_2$.\\
Row 3 implies $B_2+H_3=B_3+H_2=E_3$.\\

Thus,
\begin{eqnarray*}
I' & = & [e_1(B_1,C_2)+e_2(B_2;C_1)]Y^{E_1}   \\
   & + & [e_1(B_1;C_3)+e_3(B_3;C_1)]Y^{E_2}   \\
   & + & [e_2(B_2;C_3)+e_3(B_3,C_2)]Y^{E_3}=0 \\
\end{eqnarray*}
and $(B_k;C_j)$ are given by (\ref{4.6}) $(j,k=1,2,3)$. \\
To find $I_3(\theta)$ of $S_3(\theta)$, we have $I_2(\theta)$ is as in (\ref{3.12}) and we solve for $B_3$ by using the linear system:
\begin{tabbing}
\marca
$ \dis \left\{ \begin{array}{l}
(B_3,C_3)=0             \\
(B_3,C_1)=(H_3-H_1,C_1) \\ 
\end{array}\right. $ \> \parbox{15mm}{\begin{theo}\label{4.8}\end{theo}} \\
\end{tabbing}
Yielding: $B_3$ in the same form as $B_1,B_2$ e.g.:
\[ B_3=\frac{(H_3-H_1,C_1)}{d_{13}}(c_{23},-c_{13}),\;\;\;\;d_{13}=\left| \left( \begin{array}{cc} c_{11} & c_{13} \\
        c_{21} & c_{23} \end{array}\right) \right| \]
$e_1,e_2$ are as in $I_2$ given by (\ref{3.12})

Setting the coefficient of $Y^{E_2}$ to zero we get 
\begin{tabbing}
\marca
$ \dis e_3=-e_1\frac{(B_1;C_3)}{(B_3;C_1)}=-e_1\frac{(H_1-H_3;C_3)}{(H_3-H_1;C_1)} $ \> \parbox{15mm}{\begin{theo}\label{4.9}\end{theo}} \\
\end{tabbing}
We now must show that 
\begin{tabbing}
\marca
$ \dis \left| \begin{array}{ccc}
 (B_1;C_2)  & (B_2;C_1) &     0     \\
 (B_1;C_3)  &     0     & (B_3;C_1) \\
     0      & (B_2;C_3) & (B_3;C_2) \\
\end{array} \right| =0 $ \> \parbox{17mm}{\begin{theo}\label{4.10}\end{theo}}\\
\end{tabbing}
Now,
\[ (B_1,C_3)=l_1(B_1,C_1)+l_2(B_1,C_2)=l_2(B_1,C_2) \]
\[ (B_2,C_3)=l_1(B_2,C_1)+l_2(B_2,C_2)=l_1(B_2,C_1) \]
Thus (\ref{4.10}) becomes:
\begin{tabbing}
\marca
$ \dis (B_1,C_2)(B_2,C_1) \left| \left(
 \begin{array}{ccc}
  1   &  1  &     0     \\
 l_2  &  0  & (B_3;C_1) \\
  0   & l_1 & (B_3;C_2) \\
\end{array}
\right) \right| $ \> \parbox{17mm}{\begin{theo}\label{4.11}\end{theo}} \\
\end{tabbing}
\[ =(B_1,C_2)(B_2,C_1)[-l_1(B_3,C_1)-l_2(B_3,C_2)] \]
\[ =(B_1,C_2)(B_2,C_1)[-(B_3;C_3)]=0\;\mbox{ by $i$ of (\ref{4.6}).} \]

Setting $e_1=1$ (in $I_2(\theta)$), we get:
\[ e_2=\frac{(H_1-H_2;C_2)}{H_1-H_2;C_1)},\;\;e_3=\frac{(H_1-H_3;C_3)}{(H_1-H_3;C_1)} \mbox{ by (\ref{4.9})} \]

Thus:
\[ I_3(\theta)=I_2(\theta)+\frac{(H_1-H_3,C_3)}{(H_1-H_3,C_1)}(y_1^{c_{23}}y_2^{-c_{13}})^{\frac{(H_3-H_1;C_1)}{d_{13}}} \]
where $I_2(\theta)$ is the first integral given by (\ref{3.12}), multiplied by $\frac{1}{(H_1-H_2;C_1)}$.

Let $S_2(\bar{\theta}),I_2(\bar{\theta}$ be as in (\ref{3.1}), (\ref{3.2}) respectively. Then
\[ C_3=\left( \begin{array}{c} l_1 \\ -l_2 \end{array} \right) \;\;\mbox{ and} \]
\[ (H_3;C_3)=-l_1+l_2\;\;\;\;\;\;\mbox{ which implies} \]
\begin{tabbing}
\marca
$ \dis l_1(h_{31}+1)=l_2(h_{32}+1) $ \> \parbox{17mm}{\begin{theo}\label{4.12}\end{theo}} \\
\end{tabbing}
\[ \mbox{and }\;\;\;B_3=B_1+H_3-H_1=(h_{31}+1,h_{32}+1)\;\;\mbox{ subject to (\ref{4.12}).} \]
\[ e_3=\frac{(H_3-H_1;C_3)}{(H_3-H_1;C_1)}=\frac{2l_2}{h_{31}+1} \]
Thus
\begin{tabbing}
\marca
$ \dis I_3(\bar{\theta})=I_2(\bar{\theta})+e_3Y^{B_3}=y_2^2+y_1^2+\frac{2l_2}{h_{31}+1}y_1^{h_{31}+1}y_2^{h_{32}+1} $ \> \parbox{17mm}{\begin{theo}\label{4.13}\end{theo}} \\
\end{tabbing}
is a first integral of 
\[ S_3(\bar{\theta})=S_2(\bar{\theta})+C_3Y^{H_3} \]
\begin{tabbing}
\marca
$ \begin{array}{cccl} 
S_3(\bar{\theta}): & y'_1 & = & y_2+l_1y_1^{h_{31}+1}y_2^{h_{32}}  \\
                   & y'_2 & = & -y_1-l_2y_1^{h_{31}}y_2^{h_{32}+1} \\
\end{array} $ \> \parbox{17mm}{\begin{theo}\label{4.14}\end{theo}} \\
\end{tabbing}
subject to (\ref{4.12}).\\
Case 2. One of $l_1,l_2$ is zero, say $l_1$, then:
\[ C_3=l_2C_2 \]
Let $(H_3,C_3)$ be subject to the condition:
\[ (H_3,C_3)=l_2(H_2,C_2)\;\;\;\mbox{ which implies:} \]
\begin{tabbing}
\marca
$ \dis (H_3-H_2,C_2)=0 $ \> \parbox{17mm}{\begin{theo}\label{4.15}\end{theo}} \\
\end{tabbing}
In addition, let $S_3(\theta)$ satisfy the following inequalities:
\begin{tabbing}
\marca
$ \dis (H_3-H_1;C_1,C_3)\neq0 $ \> \parbox{17mm}{\begin{theo}\label{4.16}\end{theo}} \\
\end{tabbing}
Let $B_1,B_2$ be as in (\ref{3.12}) then:
\[ (B_1,C_2)\neq0\;\;\mbox{ implies }\;\;(B_1,C_3)\neq0 \]
\[ (B_2,C_2)=0\;\;\mbox{ implies }\;\;(B_2,C_3)=0 \]
Let $B_3=B_1+H_3-H_1$ then
\[ \begin{array}{lcl}
(B_3;C_1) & = & (B_1+H_3-H_1;C_1)\\
          & = & (B_1;C_1)+(H_3-H_1;C_1)=0+(H_3-H_1,C_1)\neq0\\
(B_3,C_2) & = & (B_1+H_3-H_1,C_2)\\
          & = & (B_1,C_2)+(H_3-H_1;C_2)\\
          & = & (H_1-H_2;C_2)+(H_3-H_1;C_2)\\
          & = & (H_3-H_2;C_2)=0
\end{array} \]
by (\ref{4.15}) and
\[ (B_3;C_3)=(B_3,C_2)=0 \]
Thus
\begin{tabbing}
\marca
$ \dis \begin{array}{rlll}
I'_3(\theta)= & I'_2(\theta)+e_3(Y^{B_3})'        &  &                     \\
            = & [e_1(B_1;C_2)+e_2(B_2;C_1)]Y^{E_1}&\;& E_1=B_1+H_2=B_2+H_1 \\
            + & [e_1(B_1;C_3)+e_3(B_3;C_1)]Y^{E_2}&\;& E_2=B_1+H_3=B_3+H_1 \\
            = & 0                                 &  &                     \\
\end{array} $ \> \parbox{17mm}{\begin{theo}\label{4.17}\end{theo}} \\
\end{tabbing}
All the relation between $B_1,B_2,B_3,C_1,C_2,C_3,H_1,H_2,H_3$ are summarized by the 2$\times$3 array:
\[ {\mathbf A}=\begin{array}{cc} 
\begin{array}{ccc} B_1 & B_2 & B_3 \end{array} &  \\
\left( \begin{array}{ccc} H_2 & H_1 & \mbox{\O} \\
                          H_3 & \mbox{\O} & H_1 \\
                           \end{array}\right) &
\begin{array} {c} E_1 \\ E_2 \end{array} 
\end{array} \]
$B_1,B_2$ are as in $I_2(\theta)$ and $B_3=B_1+H_3-H_1$ may, also, be obtained by solving the linear system:
\[ (B_3,C_3)=0=l_2(B_3,C_2) \]
\[ (B_3,C_1)=(H_3-H_1;C_1)\;\;\;\mbox{ which yields} \]
\[ B_3=\frac{(H_3-H_1;C_1)}{d_{12}}(c_{22}-c_{12}) \]
The vanishing of the coefficient of $Y^{E_2}$ yields:
\[ e_3=-e_1\frac{(B_1,C_3)}{(B_3,C_1)}=\frac{-e_1l_2(B_1;C_2)}{(B_3,C_1)} \]
setting $e_1=1$ we get
\[ e_3=-l_2\frac{(H_1-H_2,C_2)}{(H_3-H_1,C_1)}=\frac{l_2(H_2-H_1,C_2)}{(H_3-H_1;C_1)} \] 
Thus, if $S_3(\theta)=S_2(\theta)+C_3Y^{H_3}$, where $S_2(\theta)$ is as in (\ref{3.9}), $C_3=l_2C_2$ and $H_3$ is subject to the condition:
\[ (H_3-H_2,C_2)=0 \]
and the inequalities $(H_3-H_1;C_1,C_3)\neq0$, then $S_3(\theta)$ has the first integral:
\[ I_3(\theta)=I_2(\theta)+e_3Y^{B_3}=I_2(\theta)+l_2\frac{(H_2-H_1,C_2)}{(H_3-H_1,C_1)}(y_1^{c_{22}}y_2^{-c_{12}})^{\frac{(H_3-H_1;C_1)}{d_{12}}} \]
where $I_2(\theta)$ is given by (\ref{3.12}).

\section{}
\vspace{14pt} 
Example 3.
\begin{tabbing}
\marca
$ \dis y''-2y'^2+3y^2=0$ \> \parbox{15mm}{\begin{theo}\label{5.1}\end{theo}} \\
\end{tabbing}
has a first integral
\[I=-4y+\ln\left(1-\frac{16}{3}y'^2+8y^2+4y\right)\]
Let $y=y_1,\;\;y'=y_2$, then the multinomial vector form for (\ref{5.1}) is:
\[y'=y(C_1Y^{H_1}+C_2Y^{H_2}+C_3Y^{H_3}),\mbox{ where: }\]
\[H_1=(-1,1),\;\;H_2=(0,1),\;\;H_3=(2,-1)\]
\[C_1=\left( \begin{array}{c} 1 \\ 0 \end{array}\right),\;\;C_2=\left( \begin{array}{c} 0 \\ 2 \end{array}\right),\;\;C_3=\left( \begin{array}{c} 0 \\ -3 \end{array}\right).\]
\vspace{2mm}
\[I=e_1Y^{B_1}+\ln\left(1+e_2Y^{B_2}+e_3Y^{B_3}+e_4Y^{B_4}\right),\mbox{ where: }\]
\[B_1=(1,0),\;\;B_2=(0,2),\;\;B_3=(2,0),\;\;B_4=(1,0)\]
\[e_1=-4,\;\;\;e_2=-\frac{16}{3},\;\;\;e_3=8,\;\;\;e_4=4. \]
Now,
\[I'=e_1(Y^{B_1})'+\frac{e_2(y^{B_2})'+e_3(Y^{B_3})'+e_4(Y^{B_4})'}{1+e_2Y^{B_2}+e_3Y^{B_3}+e_4(Y^{B_4})}=0\]
Clearing of fractions and using the relations:
\[(B_2;C_1)=0,\;\;(B_1,B_3,B_4;C_j)=0 \; \; \mbox{ if }j\neq1,\]
yields:
\begin{tabbing}
\marca
$\begin{array}{lcl} I' & = & e_1(B_1;C_1)Y^{B_1+H_1}[1+\sum^4_{k=2}e_kY^{B_k}]\\
                       & + & e_2[(B_2;C_2)Y^{B_2+H_2}+(B_2;C_3)Y^{B_2+H_3}]\\
                       & + & e_3(B_3;C_1)Y^{B_3+H_1}+e_4(B_4;C_1)Y^{B_4+H_1}=0\end{array}$\> \parbox{15mm}{\begin{theo}\label{5.2}\end{theo}}\\
\end{tabbing}
Grouping all the coefficients of the same vector together, we get:
\begin{tabbing}
\marca
$\begin{array}{lcl} I' & = & [e_1(B_1;C_1)+e_4(B_4;C_1)]Y^{E_1}\\
                       & + & [e_2(B_2;C_2)+e_1e_2(B_1;C_1)]Y^{E_2}\\
                       & + & [e_2(B_2;C_3)+e_1e_3(B_1;C_1)]Y^{E_3}\\
                       & + & [e_3(B_3,C_1)+e_1e_4(B_1;C_1)]Y^{E_4}=0,\end{array}$\> \parbox{15mm}{\begin{theo}\label{5.3}\end{theo}}\\
\end{tabbing} 
where:
\begin{tabbing}
\marca
$\begin{array}{l}  
i)\;\;\;E_1=B_1+H_1=B_4+H_1=(0,1)\\
ii)\;\;\;E_2=B-2+H_2=B_1+B_2+H_1=(0,3)\\
iii)\;\;\;E_3=B_2+H_3=B_1+B_3+H_1=(2,1)\\
iv)\;\;\;E_4=B_3+H_1=B_1+B_4+H_1=(1,1)\end{array}$\> \parbox{15mm}{\begin{theo}\label{5.4}\end{theo}}\\
\end{tabbing}
The integral array ${\mathbf A}$ of I is:
\[  {\mathbf A}=
 \begin{array}{cc}
 \begin{array}{ccccccc}
    B_1 &  B_2  &  B_3 & B_4 & B_1+\bar{B}_2 & B_1+\bar{B}_3 & B_1+\bar{B}_4   \\
 \end{array}                          &    \\
\! \! \left( \begin{array}{ccccccccccccccc}
   H_1   & \mbox{\O}& \mbox{\O}&    H_1   & & \mbox{\O}& & & & \mbox{\O}& & & & \mbox{\O}& \\ 
\mbox{\O}&   H_2    & \mbox{\O}& \mbox{\O}& &    H_1   & & & & \mbox{\O}& & & & \mbox{\O}& \\
\mbox{\O}&   H_3    & \mbox{\O}& \mbox{\O}& & \mbox{\O}& & & &   H_1    & & & & \mbox{\O}& \\
\mbox{\O}& \mbox{\O}&    H_1   & \mbox{\O}& & \mbox{\O}& & & & \mbox{\O}& & & &    H_1   & \\
\end{array}\right)                    & \begin{array}{c} 
                                        E_1 \\ 
                                        E_2 \\ 
                                        E_3 \\
                                        E_4 \\
                                        \end{array}  
\end{array} \]

The bars over $B_2,B_3,B_4$ in columns 5,6,7 are used to indicate that they do not appear in the coefficients of $Y^{E_i}(i=2,3,4)$.
 \begin{tabbing}
\marca
$ \dis  M({\mathbf A})= \left( \begin{array}{ccccccc}
(B_1;C_1) &     0     &     0     & (B_4;C_1) &     0     &     0     &     0    \\
    0     & (B_2;C_2) &     0     &     0     & (B_1,C_1) &     0     &     0    \\
    0     & (B_2;C_3) &     0     &     0     &     0     & (B_1;C_1) &     0    \\
    0     &     0     & (B_3;C_1) &     0     &     0     &     0     & (B_1;C_1)\\ 
\end{array}\right) $  
\end{tabbing}
\begin{tabbing}
\marca
$ \dis \left\{ \begin{array}{ll}
i)\mbox{ of (\ref{5.4}) implies } &  B_1=B_4=(1,0)\\
ii)\mbox{ of (\ref{5.4}) implies } &  B_1=B_4=H_2-H_1=(1,0)\\
iii)\mbox { and } iv) \mbox{ imply } &  B_2=3H_2-2H_1-H_3=(0,2)\\
                                     &  B_3=2B_1=2(H_2-H_1)=(2,0)
\end{array}\right.$ \> \parbox{15mm}{\begin{theo}\label{5.5}\end{theo}} \\
\end{tabbing}
Setting the coefficients of $Y^{E_i}=0$, in (\ref{5.3}), $(i=1,\ldots,4)$ we get:
\begin{tabbing}
\marca
$ \dis \left\{ \begin{array}{ll}
e_4=-e_1, &  e_1=-\frac{(B_2;C_2)}{(B_1;C_1)}=-4 \\
e_3=-e_1e_4\frac{(B_1;C_1)}{(B_3;C_1)}=8, & e_2=-e_1e_3\frac{(B_1;C_1)}{(B_2;C_3)}=\frac{-16}{3} \end{array}\right.$ \> \parbox{15mm}{\begin{theo}\label{5.6}\end{theo}} \\
\end{tabbing}

We now use the $4\times7$ array ${\mathbf A}$ to find the conditions that $S(\Theta)$ has to satisfy, to have a first integral
\[I(\Theta)=e_1Y^{B_1(\Theta)}+\ln\left(1+e_2Y^{B_2(\Theta)}+e_3Y^{B_3(\Theta)}+e_4Y^{B_4(\Theta)}\right)\]
where $S(\Theta)$ is given by:
\[S(\Theta):\;y'=y[\left( \begin{array}{c} c_{11} \\ c_{21} \end{array}\right)Y^{(h_{11},h_{12})}+\left( \begin{array}{c} c_{12} \\ c_{22} \end{array}\right)Y^{(h_{21},h_{22})}+\left( \begin{array}{c} c_{13} \\ c_{23} \end{array}\right)Y^{(h_{31},h_{32})}]\]
Using $ii)$ of (\ref{5.5}) we have $B_1=B_4=H_2-H_1$ and looking at the array ${\mathbf A}$, we see that $(B_1,B_4;C_j)=0$ if and only if $j\neq1$. Thus we have
\begin{tabbing}
\marca
$ \dis \left\{  \begin{array}{ll}
a)\;\;(H_2-H_1,C_j)=0 & \mbox{ if and only if }j\neq1.\mbox{ Similarly, }  \\
&  \\
b)\;\;(B_2;C_1)=0 & \mbox{ implies }(3H_2-2H_1-H_3;C_1)=0   \\
&  \\
c)\;\;C_3=mC_2. & \mbox{ For,} (H_2-H_1;C_2,C_3)=0 \mbox{ implies }H_1=H_2, \\ 
                & \mbox{ if } C_2,C_3\mbox{ are linearly independent,}\\
                & \mbox{ since, (the order of }S(\Theta)\mbox{ is two.)} 
\end{array}\right.$ \end{tabbing}

Thus $S(\Theta)$ is reduced, by these conditions to a 9 parameter system.
\[ I(\Theta)=e_1(\Theta)Y^{B_1(\Theta)}+\ln\left(1+\sum^{4}_{k=2}e_k(\Theta)Y^{B_2(\Theta)}\right)\]
and from (\ref{5.5}), (\ref{5.6}) we can find $B_k(\Theta),e_k(\Theta)(k=1,\ldots,4)$.
In the special case when $S(\Theta)$ comes from a second order differential equation, we have the additional conditions:
\[H_1=(-1,1),\;\;\;C_1=\left( \begin{array}{c} 1\\0 \end{array}\right),\;\;C_2=\left( \begin{array}{c} 0 \\ c_{22} \end{array}\right),\;\;C_3=\left( \begin{array}{c} 0\\c_{23} \end{array}\right).\]
\[S(\Theta):\;\;\;\;y'=y[\left( \begin{array}{c} 1\\0 \end{array}\right)Y^{(-1,1)}+\left( \begin{array}{c} 0 \\c_{22} \end{array}\right)Y^{(h_{21},1)}+\left( \begin{array}{c} 0\\c_{23} \end{array}\right)Y^{(3h_{21}+2,h_{32})}]\]
writing $S(\Theta)$ as a second order differential equation we get:
\begin{tabbing}
\marca
\parbox{110mm}{$y''=c_{22}y^{h_{21}}(y')^2+c_{32}y^{3h_{21}+2}(y')^{h_{32}+1}$}
\> \parbox{15mm}{\begin{theo}\label{5.7}\end{theo}}\\
\end{tabbing}
\[ I(\Theta)=e_1y^{(h_{21}+1)}+\ln\left(1+e_2(y')^{(1-h_{32})}+e_3y^{2(h_{21}+1)}+e_4y^{(h_{21}+1)}\right) \]
where
\[ \left\{ \begin{array}{ll}
e_1=\frac{(h_{32}-1)c_{22}}{h_{21}+1}, & e_2=\frac{(h_{32}-1)^2(c^3_{22})}{2(h_{21}+1)^2c_{32}} \\
e_3=\frac{(h_{32}-1)^2c^2_{22}}{2(h_{21}+1)^2}, & e_4=\frac{(1-h_{32})c_{22}}{h_{21}+1}
 \end{array}\right.
\]

Setting $h_{21}=0,\;h_{32}=-1,\;c_{22}=2,\;c_{32}=-3$ in (\ref{5.7}) and its first integral $I(\Theta)$, we get (\ref{5.1}) and its first integral $I$.

The differential equation (\ref{5.7}) and its first integral $I(\Theta)$, is, actually, a special case of the following:
\begin{tabbing}
\marca
\parbox{110mm}{$y''=c_{22}y^{h_{21}}(y')^2+c_{23}y^{\alpha}(y')^{\beta}$}\> \parbox{15mm}{\begin{theo}\label{5.8}\end{theo}}
\end{tabbing}
where $\alpha=(q-1)h_{21}+q-2,\;\beta=h_{32}+1$.

For any integer $q\geq3$, (\ref{5.8}) has a first integral:
\begin{tabbing}
\hspace{125mm} \= \kill
\parbox{115mm}{$ \dis I=e_1y^{(h_{21}+1)}+\ln\left(1+e_2y'^{(1-h_{22})}+\sum^q_{k=3}e_ky^{(q-k+1)(h_{31}+1)}\right)$}\> \parbox{10mm}{\begin{theo}\label{5.9}\end{theo}}\\
\end{tabbing}
Setting $q=4$ in (\ref{5.8}), (\ref{5.9}), one gets (\ref{5.7}) and its first integral $I(\Theta)$.

\section{ The $p\times q$ Array (the algebraic case).}

In the following we describe the role that the general $p\times q$ array plays in determing the necessary and sufficient conditions, that the system $S$ given by (\ref{1.3}), must satisfy, for the system to have an  algebraic first integral given by (\ref{1.4}).

Let $I=\sum^q_{k=1}e_kY^{B_k}$ be a first integral of the multinomial system given by (\ref{1.3}). Making use of the derivative formula $i)$ of (\ref{2.1}) we get:
\begin{tabbing}
\marca
$ \dis \begin{array}{lcl}
I' & = & \sum^q_{k=1}e_kY^{B_k}\sum^r_{j=1}(B_k;C_j)Y^{H_j}\\
   & = & \sum^q_{k=1}(\sum^r_{j=1}e_k(B_k;C_j)Y^{B_k+H_j})=0
\end{array}$
\end{tabbing}

In the set of $q_r$ vectors ${B_k+H_j}$, leave out all vectors such that $(B_k;C_j)=0$. Let the remaining vectors form a set of $p$ distinct vectors $E_1,\ldots,E_p$, then:
\begin{tabbing}
\marca
$ \dis I'=\sum^p_{i=1}[\sum_{B_k+H_{\alpha_i}=E_i}e_k(B_k;C_{\alpha_i})]Y^{E_i}=0 $
\> \parbox{15mm}{\begin{theo}\label{6.1}\end{theo}} \\
\end{tabbing}
\begin{tabbing}
\marca
$ \dis (B_k;C_{\alpha_i})\neq0 $
\end{tabbing}
Since $E_i\;\;(i=1,\ldots,p)$ are distinct, equation (\ref{6.1}) yields a system of $p$ homogeneous linear equations:
\begin{tabbing}
\marca
$ \dis \sum^q_{k=1}e_k(B_k;C_{\alpha_i})=0\;\;\;\;\;(i=1,\ldots,p) $
\> \parbox{15mm}{\begin{theo}\label{6.2}\end{theo}} \\
\end{tabbing}
which $e_1,\ldots,e_q$ must satisfy.
The $p\times q$ array ${\mathbf A}$ is a pictorial representation of (\ref{6.1}) and is defined as follows: 
\[ {\mathbf A}=(A_{ik})\;\;\;\;\;(i=1,\ldots,p;k=1,\ldots,q), \mbox{ where: }\]
\begin{tabbing}
\marca
$ \dis A_{ik}=\left\{ \begin{array}{l}
H_{\alpha} \mbox{ if } B_k+H_{\alpha}=E_i \mbox{ and }(B_k;C_{\alpha})\neq0 \\
 \mbox{\O} \mbox{ if no such }  H_{\alpha} \mbox{ exists}
\end{array}\right. $ \> \parbox{17mm}{\begin{theo}\label{6.3}\end{theo}} \\
\end{tabbing}
The symbol $\mbox{\O}$ stands for the empty spot. 

Definition(\ref{6.3}) implies:
\begin{tabbing}
\hspace{5mm} \= \hspace{115mm} \= \kill
\> \( \dis \left. \begin{array}{l}
\parbox{110mm}{$i)$ $H_{\alpha}$ appears in the $k^{th}$ column of ${\mathbf A}$ if and only if $(B_k;C_{\alpha})\neq0$ \\
\\
$ii)$ if $H_{\alpha},H_{\beta}$ appears in the same row in columns $j,k$, respectively, then:\\
$B_j+H_{\alpha}=B_k+H_{\beta}$} \end{array} \right\} \) \> \parbox{17mm}{\begin{theo}\label{6.4}\end{theo}} \\
\\ 
\end{tabbing}
We call columns: $j,k$ linked if $ii)$ of (\ref{6.4}) is satisfied and we set 
\[L_{jk}=H_{\alpha}-H_{\beta}\]
We call $j,k$ of ${\mathbf A}$ connected if there exist  columns. $P_1,\ldots,P_s$ such that $j=P_1,k=P_s$ and columns $P_{\alpha},P_{\alpha+1}\;\;(\alpha=1,\ldots,s-1)$ are linked, and we set
\[ L_{jk}=\sum^{s-1}_{\alpha=1}L_{P_{\alpha}P_{\alpha+1}}\]
e.g. In the array:
\[ {\mathbf A}=
\left( \begin{array}{ccc}
H_u & H_v &  \mbox{\O} \\
 \mbox{\O} & H_w & H_v 
\end{array} \right) \]
columns 1,2 are linked and $L_{12}=H_u-H_v$; columns 2,3 are linked and $L_{23}=H_w-H_v$; columns 1,3 are connected and $L_{13}=L_{12}+L_{23}=H_u+H_w-2H_v$.

Note, that when the array ${\mathbf A}$ is connected, $L_{jk}$ is defined and 
\[L_{jk}=B_k-B_j \mbox{ for all }1\leq j<k\leq q\]
This implies that the difference of any 2 exponent vectors of $I$ is a linear combination of the exponent vectors of the system $S$.

It is not difficult to show that when ${\mathbf A}$ is not connected (i.e.when there exist at least two column of  ${\mathbf A}$ which can not be connected) then  ${\mathbf A}$ is the array of an integral $I$ of $S$ such that $I=I_1+I_2$ where $I_1$ and $I_2$ are both first integrals of $S$,[1].
From now on we shall assume that the array  ${\mathbf A}$ is connected. It follows from the definition of  ${\mathbf A}$ that interchanging columns $j,k$ of ${\mathbf A}$ is equivalent to interchanging the exponents $B_j,B_k$ of $I$ and interchanging rows $i,j$ of  ${\mathbf A}$ is equivalent to interchanging $E_i,E_j$ of (\ref{6.1}). Thus we identify all arrays that can be obtained from one another by an interchange of rows and or columns.

Along with ${\mathbf A}$ we define the $p\times q$ matrix

\[ M {\mathbf A}=(a_{ik}\;\;\;\;\;\;\;(i=1,\ldots,p;k=1,\ldots,q)\]
where
\begin{tabbing}
\marca
$\left. \begin{array}{lcl}
a_{ik}=(B_k;C_{\alpha}) & \mbox{ if } & A_{ik}=H_{\alpha}\\
a_{ik}=0                & \mbox{ if } & A_{ik}=\mbox{\O}
\end{array}\right\} $  \> \parbox{15mm}{\begin{theo}\label{6.5}\end{theo}} \\
\end{tabbing}
There are two kinds of arrays normal and abnormal. An array  ${\mathbf A}$ is called normal if for every $H_{\alpha}$ that appears in  ${\mathbf A}$, there exist at least one column of  ${\mathbf A}$ which does not contain $H_{\alpha}$. An array  ${\mathbf A}$ is called abnormal if there exists at least one $H_{\alpha}$ which appears in every column of  ${\mathbf A}$.
It is remarkable that when  ${\mathbf A}$ is normal we can compute $(B_k;C_{\alpha})$ without knowing what the $B_k$'s are. For, let $H_{\alpha}$ fail to appear in column $j$ of ${\mathbf A}$ then, for any $k\neq j$ we have 
\begin{tabbing}
\marca
$ \dis \begin{array} {lll}
(B_k;C_{\alpha}) & = & (B_k-B_j;C_{\alpha})+(B_j;C_{\alpha}) \\
                 & = & (L_{jk};C_{\alpha})+0=(L_{jk},C_{\alpha})
\end{array}$ \> \parbox{15mm}{\begin{theo}\label{6.6}\end{theo}} \\
\end{tabbing}
Thus, $M({\mathbf A})$, which is the matrix of the system of homogeneous linear equations given by (\ref{6.2}), can be computed, when ${\mathbf A}$ is normal.

Let ${\mathbf A}$ be an array of the exponent vectors of a system $S$, given by (\ref{1.3}). The following is a set of necessary and sufficient conditions that ${\mathbf A}$ must sotisfy for ${\mathbf A}$ to be an integral array (i.e. there exists a first integral $I$ of $S$ such that ${\mathbf A}$ satisfies (\ref{6.3})).
\begin{tabbing}
\hspace{5mm} \= \hspace{115mm}  \= \kill
\> $ \left. \begin{array}{l}
\begin{minipage}{110mm}{a) Each row of ${\mathbf A}$ must contain at least two distinct $H_{\alpha}$'s of $S$ and no $H_{\alpha}$ may appear more than once in any row or column of ${\mathbf A}$. \\
\\
b) The $L_{jk}\;\;(1\leq j< k\leq q)$ are well defined and do not equal $\bar{0}$ ($\bar{0}=$ the zero vector). \\
\\
c) If $H_{\alpha}$ fails to appear in columns $j,k$ of ${\mathbf A}$, then $(L_{jk};C_{\alpha}=0$ \\
\\
d)If $H_{\alpha}$ appears in column $j$ but fails to appear in column $k$, then $(L_{jk};C_{\alpha})\neq0$. \\
\\
e) Rank of any $q-1$ columns of $M({\mathbf A})$ equals rank of $M({\mathbf A})=q-1$. \\
\begin{tabbing}
\hspace{3mm} \= \hspace{40mm} \= \kill
f) $B_1$ satisfies the linear system: \\
\\
\>  $(B_1;C_{\alpha})=0$ \> \parbox[t]{63mm}{if $H_{\alpha}$ fails to appear in column 1.}\\
\\
\> $(B_1;C_{\alpha})=(L_{j1};C_{\alpha})$ \> \parbox[t]{63mm}{if $H_{\alpha}$ appears in column 1, but fails to appear in column $j$ (the existence of such a $''j''$ is guaranteed by the fact that ${\mathbf A}$ is normal).} 
\end{tabbing}}
\end{minipage}
\end{array} \right\} $
\parbox{17mm}{\begin{theo}\label{6.7}\end{theo}} \\
\end{tabbing}

Condition $a)$ is necessary, since $I'=0$ implies that the coefficients of $Y^{E_i}\;\;(i=1,\ldots,p)$ must vanish, thus if only one $H_{\alpha}$ appears in row $i$ then the coefficient of $Y^{E_i}$ is $=0$ contrary to assumption. Also, 
\[H_{\alpha}=A_{ij}=A_{ik} \mbox{ implies } B_j=B_k \mbox{ and }\]
\[H_{\alpha}=A_{ik}=A_{jk} \mbox{ implies } E_i=E_j\]
condition b) is a strong restriction on $S$ e.g. if 
\[  {\mathbf A}= \left(
\begin{array}{cc}
H_1 & H_2  \\
H_3 & H_4
\end{array} \right) \]
then $L_{12}=H_1-H_2=H_3-H_4$ yields a linear relation among the exponents. Also, $L_{jk}=\bar{0}$ implies $B_j=B_k$.

Conditions $c),d)$ are implied by $L_{jk}=B_k-B_j$, thus if $H_{\alpha}$ does not appear in either column, we have $(L_{jk},C_{\alpha})=(B_k;C_{\alpha})-(B_j,C_{\alpha})=0+0=0$. Similarly if $H_{\alpha}$ appears in column $k$ but not in column $j$, $(L_{jk};C_{\alpha})=(B_k;C_{\alpha})-(B_j,C_{\alpha})=(B_k,C_{\alpha})-0\neq0$.

Condition $e)$ is the requirement that the system of homogeneous linear equations in $e_k\;\;(k=1,\ldots,q)$, given by 
\[ \sum^q_{k=1}a_{ik}e_k=0\;\;\;\;\;i=(1,\ldots,p),\]
has a solution such that $\prod^{q_n}_{k=1}e_k\neq0$.

Condition $f)$ follows from (\ref{6.3}) and (\ref{6.6}) by setting $k=1$.

We now show that conditions $a),\ldots,f)$ are also sufficient for the construction of a first integral $I=\sum^{q_n}_{k=1}e_kY^{B_k}$ of $S$ given by (\ref{1.3}).

Let ${\mathbf A}=(A_{ik}),\;M({\mathbf A})=(a_{ik})\;\;(i=1,\ldots,p\;;k=1,\ldots,q)$, where $a_{ik}=0$ if $A_{ik}=\emptyset,\;a_{ik}=(L_{jk};C_{\alpha})$ if $A_{ik}=H_{\alpha}$ and $j$ is such that $H_{\alpha}$ does not appear in column $j$ of ${\mathbf A}$. Let ${\mathbf A},\;M({\mathbf A})$ satisfy conditions $a),\ldots,f)$ then we constant a first integral $I$ as follows:
\begin{tabbing}
\marca
$\begin{array}{l} 
\parbox{110mm}{$i)$ Let $B_1$ be a solution of the linear system as in f). \\
$ii)$ Set $B_k=B_1+L_{1k}$ \\
$iii)$ Set $e_1,\ldots,e_q$ to be a solution of the homogeneous linear system:\\
$\sum^q_{k=1}a_{ik}e_k=0$, (a solution is guaranteed by $e$). \\
$iv)$ Set $I=\sum^q_{k=1}e_kY^{B_k}$.}
\end{array}  $ 
\end{tabbing}

We first show that $(B_k;C_{\alpha})=0$ if and only if $H_{\alpha}$ does not appear in the $k^{th}$ column of ${\mathbf A}$. By $i)$ this holds for $k=1$, let $k>1$. Let $H_{\alpha}$ not appear in the $k^{th}$ column of ${\mathbf A}$ and let $H_{\alpha}$ also fail to appear in the first column of ${\mathbf A}$, then 
\[(B_k;C_{\alpha})=(B_1;C_{\alpha})+(L_{1k};C_{\alpha})=0+0=0 \]

$(B_1;C_{\alpha})=0$ follows from the definition of $B_1$ given by $i),\;(L_{1k};C_{\alpha})=0$ follows from condition $c)$ that ${\mathbf A}$ satisfies.
Let $H_{\alpha}$ fail to appear in column $k$ but appear in column $1$, then $(B_1;C_{\alpha})=(L_k;C_{\alpha})$ by $i)$, so that
\[(B_k;C_{\alpha})=(B_1;C_{\alpha})+(L_{1k};C_{\alpha})=(L_{k1};C_{\alpha})+(L_{1k};C_{\alpha})=0.\]
Now, let $H_{\alpha}$ appear in column $k$ but fail to appear in column $1$, then
\[(B_k;C_{\alpha})=(B_1;C_{\alpha})+(L_{1k};C_{\alpha})=0+(L_{1k};C_{\alpha})\neq0,\]
by condition $d)$ that ${\mathbf A}$ satisfies.

Finally, if $H_{\alpha}$ appears in column $k$ and column $1$, but fails to appear in some column $j$ (such a $j$ exists because ${\mathbf A}$ is normal) then
\[(B_k;C_{\alpha})=(B_1;C_{\alpha})+(L_{1k};C_{\alpha})=(L_{j1};C_{\alpha})+(L_{1k};C_{\alpha})=(L_{jk};C_{\alpha})\neq0\;,\]
by condition $d)$.

This proves our assertion for all $1\leq k\leq q$.
Let $H_{\alpha}=A_{ij},\;H_{\beta}=A_{ik}$, we show that 
\[B_j+H_{\alpha}=B_k+H_{\beta}=E_i\;\;\;\;(i=1,\ldots,p)\]
For,
\begin{tabbing}
\marca
$ \begin{array}{lcl}
B_k-B_j & = & B_1+L_{1k}-(B_1+L_{1j})\;\;\;\mbox{ (by $ii)$)} \\
        & = & L_{1k}+L_{j1}=L_{jk}
\end{array} $
\end{tabbing}
and $H_{\alpha}-H_{\beta}=L_{jk}$ (by definition of $L_{jk}$ when columns $j,k$ are linked)
Thus
\[H_{\alpha}-H_{\beta}=B_k-B_j\;\;\;\;\mbox{ and }\]
\[ H_{\alpha}+B_j=H_{\beta}+B_k=E_i \]
We can, now, show that 
\[I=\sum^q_{k=1}e_kY^{B_k}\]
is a first integral of $S$. For,
\[I'=\sum^q_{k=1}e_k(Y^{B_k})=\sum^q_{k=1}e_k\sum^r_{j=1}(B_k,C_j)Y^{B_k+H_j}\]
But $(B_k;C_j)=0$ for any $H_j$ which does not appear in column $k$, therefor the sum of the $(B_k,C_j)$'s is restricted to summing along the columns of ${\mathbf A}$.
If instead we sum along the rows of ${\mathbf A}$, we get:
\[I'=\sum^k_{i=1}(\sum^q_{k=1}e_k(B_k,C_j))Y^{E_i}\]
For each $H_j$ in ${\mathbf A}$, let $\alpha_j$ be the column of ${\mathbf A}$ in which $H_j$ fails to appear, then:

\begin{tabbing}
\marca
$ \begin{array}{lcl}
(B_k;C_j) & = & (B_k-B_{\alpha j};C_j)+(B_{\alpha j},C_j)\\
          & = & (L_{\alpha j,k};C_j)+0=(L_{\alpha j,k};C_j) \mbox{ as in (\ref{6.6})}\\
          & = & a_{ik} \;\;\mbox{ by (\ref{6.7})}
\end{array} $
\end{tabbing}
Thus 
\[I'=\sum^k_{i=1}(\sum^q_{k=1}e_ka_{ik})Y^{E_i}=0\]
by the choice of $e_k\;\;(k=1,\ldots,q)$ in $iii)$.

\end{document}